# Stray radiation produced in FLASH electron beams characterized by the MiniPIX Timepix3 Flex detector

C. Oancea[a], C. Bălan[b,1], J. Pivec[a], C. Granja[a], J. Jakubek[a], D. Chvatil[c], V. Olsansky[c], and V. Chiş[b]

[a] *ADVACAM*
 *U Pergamenky 12, 170 00 Prague, Czech Republic*

[b] *Faculty of Physics, Babeş–Bolyai University*
 *M. Kogălniceanu 1,40084 Cluj–Napoca, Romania*

[c] *Nuclear Physics Institute, Czech. Academy of Sciences*
 *Hlavni 130, 250 68 Rez near Prague, Czech Republic*

 *E–mail*: cristinabalancj@gmail.com

ABSTRACT: This work aims to characterize ultra–high dose rate pulses (UHDpulse) electron beams using the hybrid semiconductor pixel detector. The Timepix3 (TPX3) ASIC chip was used to measure the composition, spatial, time, and spectral characteristics of the secondary radiation fields from pulsed 15–23 MeV electron beams. The challenge is to develop a single compact detector that could extract spectrometric and dosimetric information on such high flux short–pulsed fields. For secondary beam measurements, PMMA plates of 1 and 8 cm thickness were placed in front of the electron beam, with a pulse duration of 3.5 μs. Timepix3 detectors with silicon sensors of 100 and 500 μm thickness were placed on a shifting stage allowing for data acquisition at various lateral positions to the beam axis. The use of the detector in FLEXI configuration enables suitable measurements in–situ and minimal self–shielding. Preliminary results highlight both the technique and the detector's ability to measure individual UHDpulses of electron beams delivered in short pulses. In addition, the use of the two signal chains per–pixel enables the estimation of particle flux and the scattered dose rates (DRs) at various distances from the beam core, in mixed radiation fields.

Corresponding author.

# Contents

## 1. Introduction and goals

### 1.1 FLASH electron beams

Modern oncological treatments are balancing between two aspects, tumour necrosis and a minimal impact on healthy tissue. Radiation therapy can deliver a high dose to the tumour spearing the healthy tissue located in the vicinity of the tumour [1]. Radiotherapy's current challenge is finding a balance between the tolerance of the healthy cell recovery to the variation of the dose fractionation and various treatments based on different types of particles [2, 3]. Electrons can be considered one of the pioneers of radiotherapy treatments due to their capability to treat superficial localized tumours owing to the nature of the electron–matter interactions. The physical aspects of the electron, the dose fall–out in the medium, and the biological response of the healthy cell encourage previous treatments with this type of particle to be used in the cases where the tumours are located superficially (at the skin level or a few cm deep) [4, 5]. A new approach is investigated due to the constant evolution of the technology and influence of the actual trends following cancer's curative treatments. FLASH treatment is a recent concept introduced in the field of radiotherapy, at origin being an effect that was observed at the cellular level if a few conditions of irradiation are met: i) the DR is intensely magnified ($\geq 40$ Gy/s) compared to the daily clinical performs ($\geq 0.01$ Gy/s) and ii) the time of irradiation should be very short [6–8]. Promising results for the FLASH effect were reported for electrons, very–high–energy electrons, protons or heavy ions [2, 9, 10].

### 1.2 Dosimetry of FLASH beams

To complete the radiobiology goal and achieve the FLASH effect in the clinic, the technological necessities in detectors suitable for those high currents are under development [11]. The challenge for providing a detector to be suitable for the FLASH treatments is to solve three features: i) dose–rate behaviour, ii) spatial resolution and iii) the detector's time response. Having a clear picture of those physical characteristics, a detector could be implemented in the dosimetric protocols [12]. The purpose of this work is to test and determine the feasibility of the prototype MiniPIX Timepix3 detectors to characterize scattered radiation from low doses to a very large range of doses that can be considered FLASH (up to 40 Gy/s). A method and setting parameters such as operation modes and acquisition time were tested for electron beams with energy levels similar to the ones used in clinic.

This study will present the impact that MiniPIX TPX3 detectors, with different sensors and designs, could have on the DR interpretation. For that, extensive attention is offered to the stray radiation of an ultra–high DR pulsed electron beam produced by a Microtron accelerator. Placed at different lateral positions of the beam core, the pixelated detectors used in the experiment could monitor and measure dosimetric aspects of the beam, such as absorbed dose, the flux of particles, and the time of arrival.

The preliminary experiment presented in this study could contribute to quantifying the scattered radiation at different distances from the target and estimate the effect of radiation on organs at risk located in the vicinity of a tumour. Future plans include detailed characterization of the radiation fields (including linear energy transfer spectra, identification of individual particles in a mixed radiation field, 2D deposited energy) in a water-phantom in therapeutical electron beams.

A hybrid pixel detector based on the ASIC semiconductor technology from the Medipix family is used to measure and characterize the stray radiation produced by FLASH electron beams [13]. TPX3 is an active detector with a remarkable wide use in different domains, such as radiotherapy (particle imaging, type of particle identification, linear–energy transfer and dose rate (DR) measurements), spectroscopic X–ray imaging, Radon monitoring, or distribution of the radiation that is found in aeronautical science [13–17]. The futuristic design for TPX3 highlights a specific characteristic of the new pixel detectors generation, the data–driven architecture: every event/pixel is sent to the integrated chip instantly [18]. Considering all the requirements that the FLASH therapy assumes, the TPX3 detector could be a great candidate for a comprehensive evaluation of the radiation field with both components, primary and secondary radiation, and also a valuable instrument for real–time monitoring of the events.

## 2. Materials and Methods

### 2.1 MiniPIX TPX3 Rigid

New dosimetric challenges are imposed by the emerging modern cancer treatment techniques and suitable detectors and dosimetry protocols need to be developed when dealing with ultra–high dose rate pulses (UHDpulse) such as ultra–short pulses of MeV–level electrons. The MiniPIX TPX3 detectors used in this study are provided by ADVACAM [19] and they were developed within the Medipix Collaboration at CERN. The ASIC read–out chip contains a matrix of 256 × 256 pixels (total 65536 independent channels, where one pixel corresponds to 55 µm) and an active sensor area of 14 mm × 14 mm[20]. The detector's sensor can be manufactured from different materials (Si, CdTe, GaAs), with variable thicknesses (from 100 µm to 2000 µm) [21]. TPX3 detector provides two per–pixel signal read–out for simultaneous registration of the energy and time at the pixel level. An illustration of the MiniPIX TPX3 detector can be seen in Fig. 1a. For this work, the thinnest available sensor, 100 µm, was selected in order to reduce the integration of particles inside the detector's chip, and the 500 µm Si sensor layer was chosen to ensure a longer path of the particles inside the chip.

## 2.2 A customized MiniPIX TPX3 Flex

The MiniPIX TPX3 Flex, Fig. 1b, is a new customized version of the MiniPIX TPX3 family, designed within the UHDpulse Project to characterize the UHDpulse electron and proton beams. In this work, the MiniPIX TPX3 detector was tested for its suitability of use in high–intensity electron beams. Tests for validating the most suitable configurations (detector's operation mode, acquisition time, threshold, per–pixel saturation, sensor's type, and thickness) were performed. Compared to the rigid version of the MiniPIX TPX3 detector, the customized Flex version has a unique pattern: the sensor was moved 5 cm away from the electronic components using a flexible cable. To maintain the external perturbation at the lowest level, the sensor was placed on an extruded graphite, see Fig. 1b.

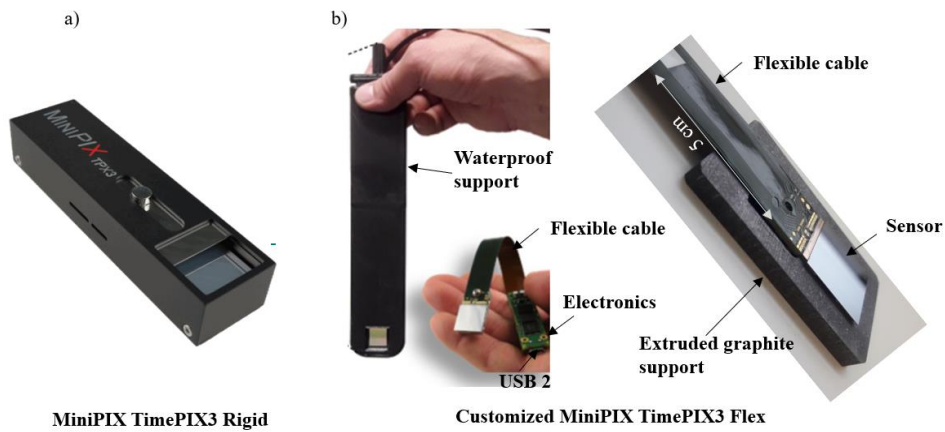

**Figure 1. a)** The MiniPIX Timepix3 detector consists of a semiconductor radiation–sensitive sensor (100 µm silicon, full size 14 mm x 14 mm) bump–bonded to the ASIC Timepix3 readout chip with an array of 256 by 256 pixels and the **b)** Customized MiniPIX Timepix3 Flex free of metal holders, screws, and cooling elements (all metallic parts were replaced by carbon and ABS plastic). The readout electronics are detached 5 cm from the chip–sensor assembly (radiation sensitive element) to minimize scattering and self–shielding.

## 2.3 Detector operation and readout modes

Timepix3 detectors operate each pixel simultaneously in time and energy mode. The time of arrival (ToA) can be registered with a resolution of 1.6 ns [22]. The Time over Threshold (ToT) registers the deposited energy with an energy resolution of a few per cent, e.g., 5 keV full width at half maximum (FWHM) for 60 keV gamma-rays, and 35 keV FWHM for 5.5 MeV $^{124}$Am alpha particles [23].

For higher fluxes, the detector can be operated in frame mode (Event + iToT) and measures the total/integrated per–pixel deposited energy (iToT, which provides spectral/dosimetry information) and the number of hits in each pixel (events/counts, which provides information regarding the beam intensity). In this study, both detectors were operated in frame mode (Event + iToT) with a 5 keV threshold and a bias recommended during calibration (50 V for the 100 µm, 120 V for the 500 µm). Data were saved as individual files in "txt" format. The dark noise signal enables measurements of low–LET particles with high precision [20]. Given the per–pixel calibration, an adjustable threshold is made in each pixel, allowing a detection efficiency close to 100% for heavy charged particles, making these detectors unique and suitable for this study. The threshold calibration and the pixel–by–pixel calibration was performed by the company ADVACAM using X–ray radiation sources of different energies [23]. When a particle reaches the detector's sensor, the charge diffuses in both horizontal and lateral directions and spreads over several pixels, creating the so–called "cluster", which is identified as a single particle track [24].

In Fig. 2a. and Fig. 2b, there are two illustrations of 2D and 3D per–pixel deposited energy by clusters produced by scattered particles that hit the detector's sensor surface (in this case 100 μm thick Si sensor).

The Timepix3 detector can be operated in two modes: **data–driven** (for particle fluxes up to $10^5$ particles/cm$^2$/s) and **frame mode** (for low and very high particle fluxes, >$10^5$ particles/cm$^2$/s).

**2.4 Microtron electron beam accelerator**

The microtron accelerator of the Nuclear Physics Institute, Czech Academy of Science (NPI CAS), Prague was used to produce monoenergetic pulsed electron beams with energy ranging from 15.7 to 23 MeV at high frequency and various beam intensities. Pulsed electron beams are extracted with a macro–pulse length of 3 μs and a 425 Hz repetition rate.

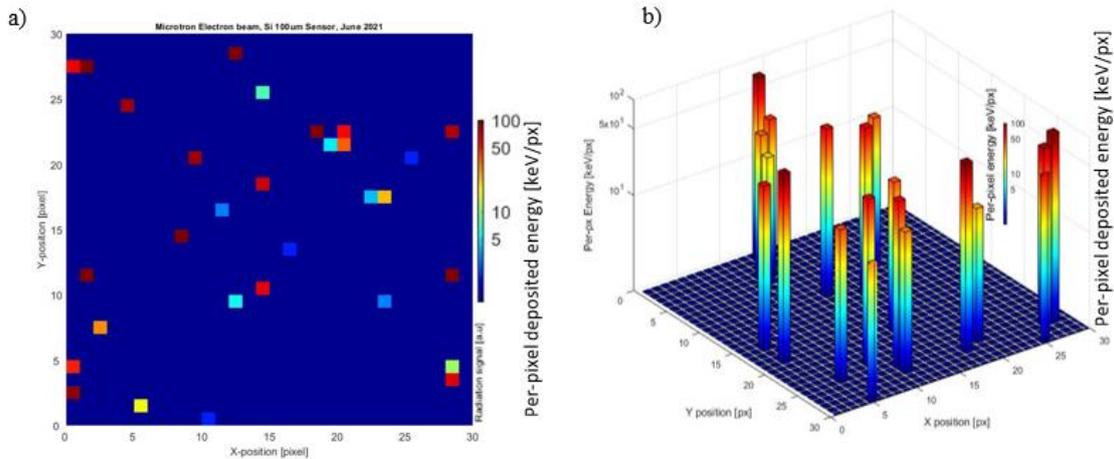

**Figure 2. a**) 2D and **b**) 3D registration of per–pixel deposited energy by 23 MeV electrons measured with a MiniPIX TPX3 Flex detector with a 100 μm Si sensor, placed perpendicular to the beam direction, at 10 cm from the beam axis. Only a part of the detector pixel matrix is shown (30 x 30 pixels = 1.65 x 1.65 mm)

## 3. Measurements

To achieve a FLASH effect (DR up to 40 Gy/s), the accelerator was used at high power, up to 1000 nA. The entire setup was placed in a shielded bunker to minimize unwanted background and to eliminate electromagnetic interference. To monitor the dose of the primary electron beam, an ionization chamber was positioned in the setup. The TPX3 detectors were placed, in turns, behind 1 cm PMMA phantom and 8 cm PMMA phantom to measure the composition, spatial, time, and spectral characteristics of the secondary radiation for a wide–range of DR intervals: from 2 Gy/s up to 40 Gy/s. In Fig. 3 a schematic representation of the experimental setup can be seen.

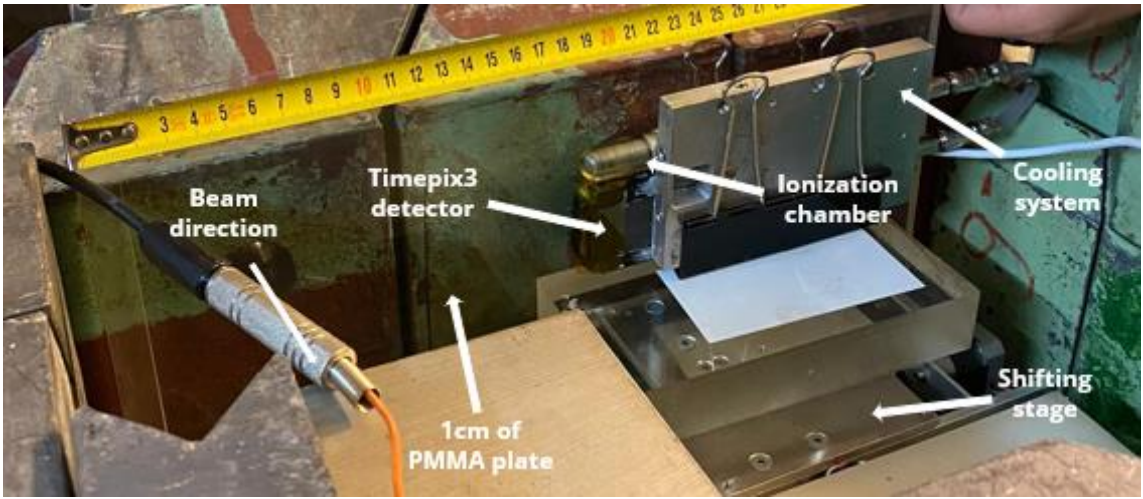

**Figure 3.** Experimental setup at the Microtron accelerator. The beam was collimated at the entrance in the bunker using a 3 cm collimator. The MiniPIX TPX3 Flex was placed laterally to the beam behind 1 cm PMMA (and 8 cm PMMA) plates and 15 cm from the incident electron beam, on a shifting stage. The setup is placed inside a massive Pb shielding to avoid large background from the accelerator.

## 4. Results and Discussion

### 4.1 Characterization of stray radiation using MiniPIX TPX3 Rigid

The particle flux and DR by a pulsed electron beam measured using the MiniPIX TPX3 detector with a 500 μm Si sensor are displayed in Fig. 4. The detector was placed behind a PMMA plate of 1 cm thickness, perpendicular to the beam, at 10 cm from the beam core. Electron beams with a nominal energy of 23 MeV and various intensities ranging from ~100 nA (4 Gy/s) up to 1000 nA (40 Gy/s) were delivered to measure the scattered particles produced laterally to the beam core.

The particle flux and the DRs were measured at 3 beam intensities to study the detector's ability to cope with UHDpulse beams of 1000 nA intensity and more. Measuring firstly at 4 Gy/s, a scattered DR of about $2 \times 10^{-5}$ Gy/s at 10 cm distance from the beam core was registered. A linear response of beam intensity versus DR was obtained, allowing us to extrapolate and estimate the scattered DRs at higher intensities than the measured ones. At 1000 nA, the incident beam becomes unstable, so the particle flux and the DRs presented as the blue dots in Fig. 4 undergo fluctuations. The detectors reading, in this case, could detect this type of fluctuation and, moreover, could display when those events occurred. Data are displayed for individual pulses for 10 seconds measurement time.

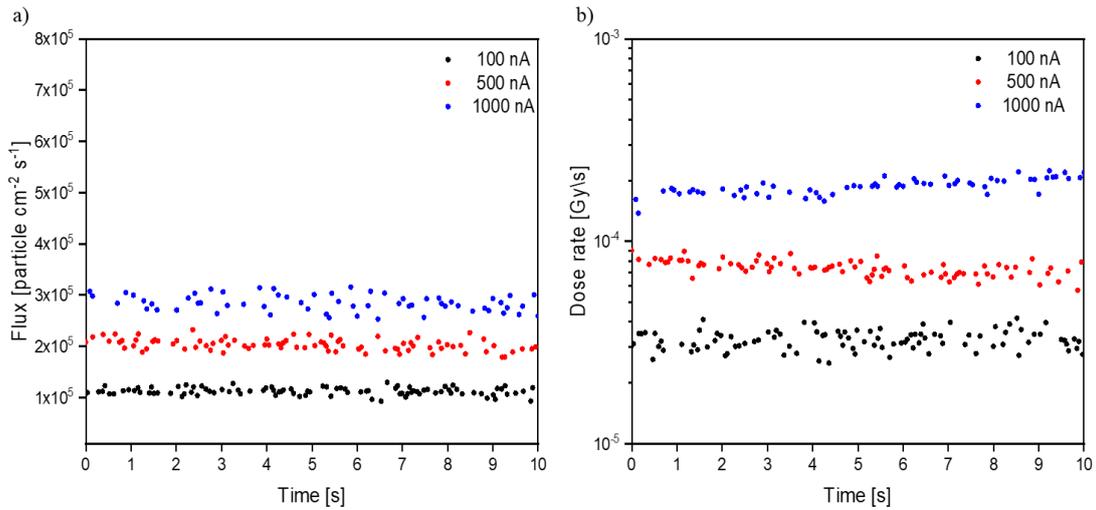

**Figure 4.** The **a)** particle flux and **b)** DRs of scattered radiation, produced by an electron beam of ~100, 500, and 1000 nA intensities were measured during 10 seconds. The background was subtracted. The MiniPIX TPX3 rigid detector with Si sensor with 500 µm thickness (bias 120 V), was placed perpendicular to the beam direction, at 10 cm lateral distance from the beam core, behind a 1 cm PMMA plate.

Observing the linear trend for both variables, the response of the MiniPIX TPX3 detector was studied at a low–intensity beam, approx. 50 nA. In this case, an 8 cm PMMA plate was placed in front of the beam to produce a higher amount of stray radiation, as Fig. 6 shows. Maintaining the intensity of the beam to approx. 50 nA, the detector was placed at 6 different lateral positions, perpendicular to the beam direction.

The flux of particles, Fig. 5a, and the DR, Fig. 5b, produced by stray radiation at every detector position sustain the same tendency during the beam on duration, 40 seconds. As expected, the particle flux and DR are increasing when approaching the beam core. In this case, the detector could be operated in data–driven mode and provide cluster information up to 6 cm distance which is a 4.5 cm distance from the border of the beam collimator. A saturation in the detector used in data–driven mode can be seen due to the fact that it reaches the upper detection limit of 2.35 million hits/s. For such high fluxes, we recommend switching the data collection mode into Frame mode (Event + iToT) and to save the files as individual files with the extension ".txt". The output would be 2 files one displaying the cumulated energy in each pixel for each pixel position and the other file displays the number of hits in each pixel for the entire sensor surface.

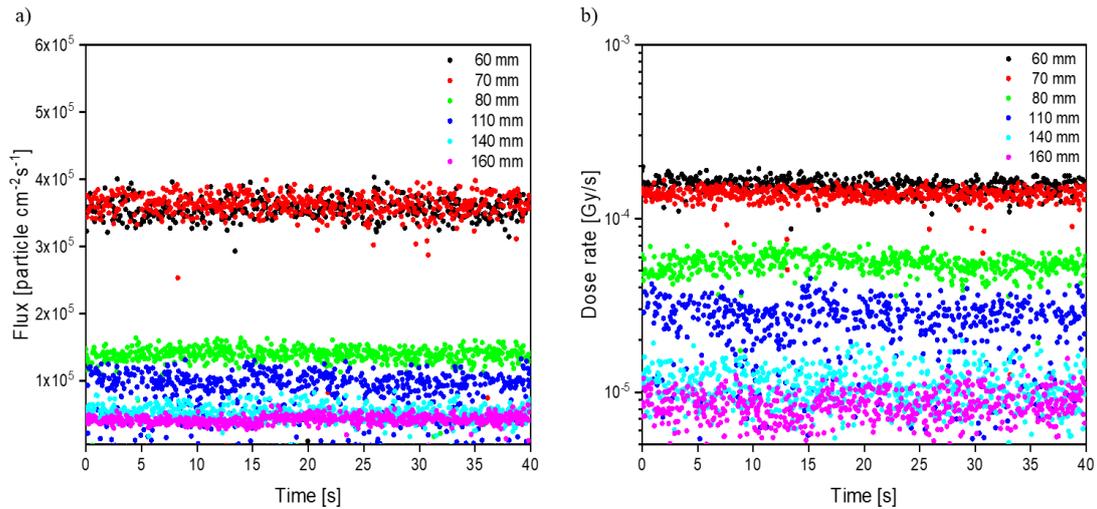

**Figure 5. a)** Particle flux and **b)** DRs of scatter radiation produced by a 23 MeV electron beam of ~50 nA. A bias of 200 V was applied to the detector and the measurements were done with the detector perpendicular to the electron beam. The MiniPIX TPX3 rigid detector with Si sensor of 500 µm thickness was placed behind an 8 cm PMMA plate, at 6, 7, 8, 11, 14, and 16 cm lateral distances from the beam axis.

### 4.2 MiniPIX TPX3 Flex in UHDpulse electron beams

The development of the new MiniPIX TPX3 Flex detector facilitated another level of analyses of these UHDpulse beams. The significant advantage of this type of detector is the opportunity to independently investigate every pulse, thus having a comprehensive characterization of the FLASH individual pulses.

The newly developed MiniPIX TPX3 Flex detectors with Si sensors of 100 and 500 µm thickness were successfully tested at the Microtron facility to characterize the stary radiation produced by electron beams with DR of up to 40 Gy/s. The detectors were placed in turns at two distances from the centre of the entry window of the beam, behind a 1 cm thickness PMMA plate. Due to very high particle fluxes, the detectors were operated in Frame mode (Event + iToT) with an acquisition time of each frame of 500 µs. The integrated energy deposited by individual pulses of electron beams is illustrated in Fig. 6 for both detectors used: MiniPIX TPX 3, 100 µm (top row) and 500 µm (bottom row) Si sensors for various beam intensities/DRs.

The integrated per–pixel deposited energy increases gradually as a result of increasing the delivered DR, see Fig. 6. Resulting in a larger number of pixel hits at the detector's surface, proportional to the particle flux. Measuring with two different thicknesses of the sensors highlights the detector's design impact since the thickness of the sensor influences the per–pixel radiation signal by increasing the sum of deposited energy in a thicker sensor and decreasing the sum of pixel hits in a thinner sensor. These consequences are related to charge sharing at the sensor level. In Fig. 6, the colour logarithmic gradient is assigned to the total deposited energy of the particles in both detectors, which increases with the delivered DR. None of the detectors reached saturation. In each frame, the mean value of the sum of integrated deposited energy and the sum of events can be seen. Considering the longer path of the particles in the MiniPIX TPX3 Flex Si 500 µm, the value of the integrated deposited energy is superior relative to the same position of the Si 100 µm detector.

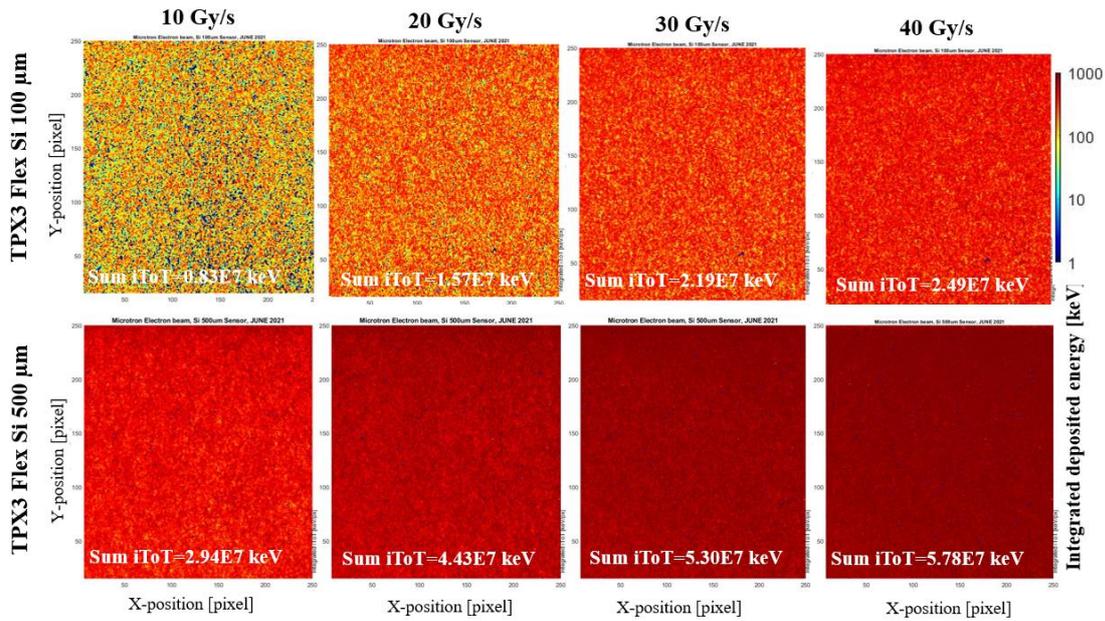

**Figure 6.** Integrated per–pixel energy measured at 15 cm distance from the beam core using the MiniPIX TPX3 Flex Si sensor of **(top row)** 100 µm and **(bottom row)** a 500 µm thickness at a pulsed-field of 15.7 MeV electrons of 10, 20, 30 and 40 Gy/s delivered DR. Each frame contains 256 x 256 pixels and represents the response of individual pulses with a length of 3.5 µs. The detector's acquisition time was set at 500 µs.

## 5. Conclusions

The preliminary results show the suitability of MiniPIX TPX3 to be used in the characterization of stray radiation produced by UHDpulse electron beams. Operating the detector in frame mode (Event + iToT) allows for measurements of integrated energy deposited by particles in the entire sensor and statistics regarding the number of events. Setting a low acquisition time, in our case 500 µs, provides information on deposited energy of individual pulses. On the other hand, when the detector was operated in data–driven mode, the UHDpulse fields could be characterized up to a 6 cm distance from the beam core. In conclusion, the customized version of MiniPIX TPX3 Flex can be further used to characterize UHDpulse beams. Furthermore, the method presented in this study can be implemented for electron beam facility commissioning or for benchmarking Monte Carlo simulations and/or treatment planning systems.

### Acknowledgments
This work was carried out within the 18HLT04 UHDpulse project from the EMPIR program co–financed by the Participating States and from the European Union's Horizon 2020 research and innovation program.